# Generative Methods for Urban Design and Rapid Solution Space Exploration

**Yue Sun***, **Timur Dogan**

*Correspondent\*: ys954@cornell.edu*
*Environmental Systems Lab, Cornell University, Ithaca, 14853, N.Y., U.S.A.*
*Cornell University, Ithaca, 14853, N.Y., U.S.A.*

**Abstract**

Rapid population growth and climate change drive urban renewal and urbanization at massive scales. New computational methods are needed to better support urban designers in developing sustainable, resilient, and livable urban environments. Urban design space exploration and multi-objective optimization of masterplans can be used to expedite planning while achieving better design outcomes by incorporating generative parametric modeling considering different stakeholder requirements and simulation-based performance feedback. However, a lack of generalizable and integrative methods for urban form generation that can be coupled with simulation and various design performance analysis constrains the extensibility of workflows. This research introduces an implementation of a tensor-field-based generative urban modeling toolkit that facilitates rapid design space exploration and multi-objective optimization by integrating with Rhino/Grasshopper ecosystem and its urban analysis and environmental performance simulation tools. Our tensor-field modeling method provides users with a generalized way to encode contextual constraints such as waterfront edges, terrain, view-axis, existing streets, landmarks, and non-geometric design inputs such as network directionality, desired densities of streets, amenities, buildings, and people as forces that modelers can weigh. This allows users to generate many, diverse urban fabric configurations that resemble real-world cities with very few model inputs. We present a case study to demonstrate the proposed framework's flexibility and applicability and show how modelers can identify design and environmental performance synergies that would be hard to find otherwise.

# Introduction

The UN predicts that rapid urbanization and population growth will lead to 2.5 billion new urban dwellers requiring a significant amount of new construction and urban renewal by 2050 (United Nations Environment Programme, 2021). Sustainable urbanization depends increasingly on the successful management of urban growth. In practice, practitioners often have to deal with various competing criteria and information from stakeholders (Scott et al., 2002) and thus modeling tasks are often open-ended, confronting modelers with a wicked problem with no single correct answer (Rittel & Webber, 1973) that translates the preferences of multiple stakeholders into a coherent consensus which is consistent with the expressed preferences of all individuals (Arrow, 1950). This increases the need for computational tools that can help practitioners to leverage generative workflows and automation and allow modelers to incorporate data driven and simulation driven performance analysis in urban design (Besserud & Hussey, 2011; Singh & Gu, 2012).

To design generative modeling frameworks, researchers have provided various definitions of urban form and offered distinct approaches to characterize typologies of urban morphologies and to describe the underlying association with urban growth processes (Gauthier & Gilliland, 2006; Kropf, 2009). Some researchers (Parish & Müller, 2001) applied L-systems algorithm to produce visually plausible results that try to imitate the street network of real cities which follow certain types of patterns on different scales. Other researchers (Raimbault & Perret, 2019) have proposed to use urban morphogenesis models, grid generator models and cellular automata models to emulate urban sprawl . Further, researchers used genetic algorithm to optimize land-use planning (Cao et al., 2011), applied agent-based modeling (Lechner et al., 2006) that can yield plausible results, and adopted an example-driven generation strategy (Nishida et al., 2016) that real network examples were extracted into street generation templates which were woven automatically. However, researchers have also noted that the simulation of urban configurations at large scale is time consuming (Lechner et al., 2006; Watson et al., 2008) and thus impacts the feasibility of devised tools in planning and urban design practice. Moreover, these distinct approaches to urban form generation require different inputs, have different constrains and remain fragmented across many different software platforms.

To address workflow fragmentation, many researchers have proposed integrated form generation and analysis pipelines to explore performance evaluation of urban design solutions (Bruno et al., 2011). Researchers (Brasebin et al., 2017) explored building configurations and regulations using a stochastic building generator and a performance analysis workflow to facilitate decision-making process of administrations, planners and citizens. Other researchers (Wortmann & Natanian, 2020) implemented multi-objective urban design exploration using a predefined array of building types and carefully limited parametric variables to keep the design space dimensionality manageable. Additionally, researchers (Miandoabchi et al., 2013) presented a framework to design road networks in a multi-objective environment, that considers street orientation, lanes allocation, and reserve capacities. While these studies explored parametric design, analysis and optimization, geometric complexity of the models had to be reduced. This limits the searchable solution space and also biases the sampling of models since their selection does not guarantee a good representation of the real variability of possible design solutions.

Hence, researchers have proposed integrated frameworks to produce more realistic, complex urban configurations. An urban modeling method using genetic algorithm for optimization has been proposed (Koenig et al., 2013, 2020). However, other researchers found limitations with the speed of convergence and stability of the underlying stochastic urban form generators (Wortmann et al., 2017) and the original authors stated that the required inputs for models can sometimes be challenging to grasp and thus may hinder design applications (Koenig et al., 2020). Further, some researchers came up an integrated workflow by using tensor field to generate street networks at urban scale and applied neural network for optimization (Chen et al., 2008; Mustafa et al., 2020). However, the study only aimed to tackle planning problems of flood-sensitive urban layout and left unexplored a lot of potential usage of tensor field and integration of other performance analytical tools.

To better support the planning or design process of different phases and devise an extensible environment to address diverse design contexts and problems, this paper proposes to extend the idea of tensor field and introduces an integrated generative design framework that quickly creates urban design variants for multi-objective design performance evaluation. The framework is implemented within Rhino Grasshopper and can be combined with other generative tools (Koenig et al., 2013, 2020) and can also be coupled with advanced analysis tools for different performance criteria such as mobility (Dogan et al., 2018), spatial accessibility (Sevtsuk & Kalvo, 2016), environmental/microclimate (Kastner & Dogan, 2018), daylight and insolation (Solemma, 2021). The advantage to employ tensor field for network generation is that it provides a level of abstraction that lets the user define the number of parametric variables and provides a way to adjust designs globally with very few inputs. Specific parameters, like street width or building height, can be defined as a spatial gradient that can controls these parameters locally and, therefore, allow users to make global urban design changes with fewer, high level input parameters. We believe this enhances design space exploration and allows modelers to sample more design variety and diversity with fewer parameters using advanced sampling strategies, like Latin Hypercube Sampling ("Latin Hypercube Sampling," 2021) to ensure that the set of models picked for analysis is a very good representative of the real design variability.

## Methodology

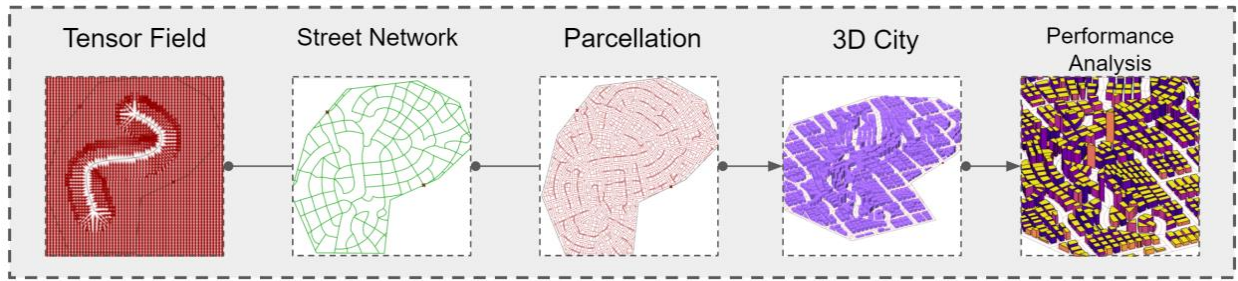

*Figure 1. The pipeline of the proposed framework. Tensor Field also encodes non-geometric information, i.e., zoning keys and building properties.*

Transportation and land use are the subsystems that have traditionally been the focus of applied urban modelling. They integrated the essential process of spatial urban development that govern how streets and buildings are configured (Wegener, 1994, 2004). However, constraints that shape these subsystems can be different in every city in a way that the importance of and the interactions between these constraints may vary. Moreover, they may be used to describe different street patterns or building typologies that can be found in different urban environments. Our framework aims to allow modelers to incorporate important aspects of these subsystems and to arbitrarily combine input data to define the driving forces for urban form generation to provide modelers with the flexibility to create a wide variety of urban morphologies. These generative driving forces are described as tensor fields by the user and serve as an abstract and generalized approach to defining input data as a spatial gradient that is sampled by the generative process. For example, modelers can use the tensor fields to encode essential urban design constraints such as terrain features like rivers or mountain slopes and design intentions, such as connections to important locations, view axis that feature landmarks, zoning targets, demographic information to drive the urban from generation. One assumption we make is that the street network must be established first to divide up the land. The street network then serves as the base geometry for parcellation and building massing generation in later modeling steps. Thus, the proposed workflow can be summarized in five main steps: 1) Incorporation of inputs and design intentions into Tensor Fields. 2) Street Network Generation. 3) Land Parcellation. 4) Building Mass Creation. 5) Performance Analysis as shown in (Figure 1).

### Overview of Tensor Field Method

The proposed system uses points, curves, and polygons to represent urban and contextual elements that influence the street network generation. For example, geographical features, like rivers or hills, often impose constraints that make it more desirable to orient streets along a river edge or terrain contour. These constraints or intentions can be encoded into a unitized tensor field (Direction Field). The proposed method then collects all relevant direction fields and executes a field tracing algorithm to create a network of curves denoting streets. The magnitudes of tensors (Magnitude Field) provide a way for users to input numerical information to control the magnitude intensities. It allows users to specify a spatial gradient that can control these parameters locally and, therefore, allows users to make global design changes with fewer inputs. As an example, Supplementary Figure S1 shows the results of imposing population density map to control the density of the generated street network.

**Tensor Field Generation**

Tensor fields are defined within a minimum bounding rectangle specified by the user. They are discretized into finite steps in X and Y direction similar to pixels in a bitmap. And each pixel is associated with a node - the position of which is determined as $N = \{ p \mid p = (x_{ij}, y_{ij}) \in R^2, i = 1,..., n, j = 1,…, m\}$, where n and m are the user-specified steps on X and Y direction. These nodes are utilized for vector tracing and for storing corresponding magnitude intensities. Since the tracing process is implemented iteratively, starting from a seed point, our algorithm factors in nearby vector directions and magnitudes to determine the next node location.

**Default tensor field** (Figure S1b Default TF)**:** In this paper, a tensor $T$ is a two-dimensional vector associated with one pixel node whose position is $p = (x_p, y_p)$. By default, all the tensors in the specified domain are $\begin{bmatrix} cos\theta \\ sin\theta \end{bmatrix}$, where $\theta \in [0, 2\pi]$ adjusts the orientation of the default tensor field map. The major eigenvector of $T$ is assigned to be $\begin{bmatrix} cos\theta \\ sin\theta \end{bmatrix}$, and the minor eigenvector to be $\begin{bmatrix} cos(\theta + \frac{\pi}{2}) \\ sin(\theta + \frac{\pi}{2}) \end{bmatrix}$, which makes the major eigenvector of the same direction as the corresponding tensor, and its minor eigenvector is perpendicular to the major one. The distinction is helpful when we use major and minor eigenvector fields by turns to create a street network with intersections.

**Tensor field with input geometry** (Figure S1b Influence Range)**:** Once we input a geometric element and specify the range of influence, tensors within that range will immediately point to their corresponding closest points $(x_c, y_c)$ on the input geometry, and can be defined as the following:

$$T(\boldsymbol{p}) = \frac{m}{\sqrt{x^2+y^2}} \begin{bmatrix} cos\theta & -sin\theta \\ sin\theta & cos\theta \end{bmatrix} \begin{bmatrix} x \\ y \end{bmatrix} \tag{1}$$

where $m$ is a user-defined magnitude of a tensor, $\theta \in [0, 2\pi)$, $x = x_p - x_c$ and $y = y_p - y_c$. Without input magnitude field, the generated field is a direction field in which the coefficient $m$ is equal to the specified length of a step. This makes the numerical value of tensors in the directional field unanimous. If a magnitude field is given, $m$ will be the corresponding numerical value at the node $p$. Hence, values may vary across one field. From equation (1) we know that $m$ is responsible for the density of a street network. When it becomes larger, the tracing step's length rises, resulting in a bigger distance between intersections, hence, a sparser network and vice versa.

**Smooth the tensor field:** the tensors outside the influence range will remain unchanged. However, we offer operation to smoothing the tensors as a user option by the following equation:

$$T(\boldsymbol{p}) = e^{-d|\boldsymbol{p}-\boldsymbol{p}_c|}T_{old}(\boldsymbol{p}) \qquad (2)$$

Where $d$ is a decay constant used to adjust the smoothness of the field, $p$ is the position of a pixel node, $p_c$ is the closest position to the pixel node on the boundary of influence range, $T_{old}(\boldsymbol{p})$ is the tensor before smoothing (Supp Fig S2b).

**Combination of tensor fields** (Supp Fig S3)**:** We may encounter scenarios with multiple design elements. This means that all these individual tensor fields need to be weighted and merged into a unified one calculated by the following equation:

$$T(\boldsymbol{p}) = \sum_i w_i e^{-d|\boldsymbol{p}-\boldsymbol{p}_c^i|}T_i(\boldsymbol{p}) \qquad (3)$$

in which $i$ denotes the sequence number of urban input element, $w$ is the weight for one individual tensor field indicating the user-specified dominance of each input geometry, $d$ is a decay constant used to adjust the smoothness of the field, $\boldsymbol{p}$ is a node, $\boldsymbol{p}_c^i$ is the closest position to the node on the corresponding geometric element, $T_i(\boldsymbol{p})$ is the original tensor at node $\boldsymbol{p}$ for the input geometry $i$. Note that $w$ and $e^{-d|\boldsymbol{p}-\boldsymbol{p}_c^i|}$ both function as the weight to determine a tensor. Such an arrangement takes the distances from nodes to input geometries into account to prevent sole dominance of certain input elements.

### Street Network Generation

Once all individual tensor fields are merged, we implement a hyperstreamline tracing algorithm (Jobard & Lefer, 1997; Zhang et al., 2007) to generate street networks. The method is a flow tracing technique where a hyperstreamline embodies a curve that is tangent to an eigenvector field everywhere along its path. A hyperstreamline either walks along the major eigenvector field or minor. The formation of the network is by the iterative tracing of major and minor hyperstreamlines. Our method is based on the works of (Chen et al., 2008; Jobard & Lefer, 1997). However, the limitation of only producing fixed number of hierarchies of streets (i.e., main, major, minor streets) decreases the flexibility of their framework (Chen et al., 2008). Our workflow allows users to generate as many levels of streets as needed by iteratively performing the same operations (Supp Fig S4).

**Tracing hyperstreamlines:** We adopt the Euler method ("Euler Method," 2021) to handle tensor fields in which an average tensor is computed at each step of tracing by extracting neighbor tensors around the corresponding point $\boldsymbol{p}$. Then the next point $\boldsymbol{p}_{next}$ can be calculated using $\boldsymbol{p}_{next} = \boldsymbol{p} + T_{average}(\boldsymbol{p})$. The tracing process stops when the hyperstreamline reaches the specified design boundary or intersects with or is too close to another existing hyperstreamline. However, to create a network of hyperstreamlines, we need to derive a seed scheme from which each hyperstreamline grows: Given an initial seed that can be specified manually or automatically, we first traverse along the major eigenvector field for creating the first hyperstreamline. A hyperstreamline encompasses multiple points. So, starting from the first point, we define that every other point along the hyperstreamline is the seed point. These seed points will be the initial points of the next hyperstreamlines that travel through the minor eigenvector field. Performing these operations iteratively will result in the first level of the street network. Additionally, the points in-between defined seeds of high-level streets will serve as seeds for the next lower level of the street network. Note that seed spacing determined by *m* influences the density of a network.

**Block Parcellation**

The generated streets can be inflated with user-specified widths and intersect to yield closed regions that serve as blocks. Our implementation uses the open-source Clipper Library (Johnson, 2014). With the resulting blocks, we implement the Oriented Bounding Box Subdivision algorithm for block subdivisions (Vanegas et al., 2012). Given one block, we first compute its minimum bounding box, then calculate the split line along short edges, split the geometry into halves, and apply the process recursively. The outputs can be tweaked by adjusting parcels' aspect ratio and area range.

In some scenarios, we want various sizes of building lots. For instance, regions with low floor area ratios may be ideal for small lots (e.g., residentials), whereas areas with high FAR may end up with large lots (e.g., office buildings). There are many ways to categorize initial parcels into several clusters before the subdivision. For instance, we can employ a density map. The idea is to sort the parcels by their areas and implement sampling considering their relative distances to the user-specified points of interest. Then the percentile of large and small parcels is specified so that, for each iteration, we can obtain a relatively stable portion of large and small ones. Such a method is used in the case study. Besides, we can also stipulate an area threshold to differentiate large and small ones (Supp Fig S5a). Each cluster has its target area, limiting the partitioning steps (Supp Fig S5b) and, consequently, reaching various building lots.

**Building Mass Generation**

The tensor field method can also be used for building mass generation. For example, we can define buildings of different heights by adding additional tensor field maps that encode land use and population density. These maps can be inputs that are sketched out in advance with stakeholders to articulate design goals or may be generated from existing data or analysis. Thus, the approach can be expanded by an arbitrary number of other potentially interesting design drivers such as environmental parameters like wind, access to sun, and allocation of urban services and population demographics.

**Density Map:** We introduce a method to generate building mass by an input population density map (PDM). Each node stores a specified population ratio value (Figure 2b, 3b). By sampling all the nodes within a given building lot and summing up the ratios, we obtain the accumulated population ratio for the lot. Subsequently, the number of people the given building lot should accommodate is computed by multiplying the ratio and the overall population of the entire design area. Since the area for each building lot is different, the numbers of nodes inside given lots are different, so their population capacities are different. Building heights corresponding to the given lots can be calculated with empirical statistics of the average area per person. Consequently, a 3D masterplan with various building heights can be achieved by extruding lots to the calculated heights. The following equation can express the process:

$$H_i = (D_i * P * U * S)/(M * A) \tag{4}$$

where $H_i$ is the height of a building, $D_i$ is the population proportion of the given lot, $P$ is the total population, $A$ is the floor area of the given lot. $M$ is the area percentage of a building used for operational purposes. $U$ stands for square meters of floor area per person, which comes from empirical data. And $S$ is the average height for one story of buildings. As the building mass is the direct outcome of the PDM, there remains a chance that the input PDM is not compatible with the network and could yield buildings that are too narrow or tall (dark blue buildings in Figure 3c). Therefore, either certain correlations between the network and the additional input should be established before the generation of streets (Figure 2a) or constraints on building heights concerning aspect ratios and areas of corresponding lots should be imposed in parallel with the network (Figure 3a). These additional constraints can prevent building outcomes that are deemed undesirable (Figure 3d). Additionally, through the modular chaining of generative components (Figures 2a, 3a), different intentions and outcomes can be modeled. If a user defines the population density, all other features, such as streets and building density, can be derived from the population density map. A user could, however, reverse the chaining order of components and start with the street network that, for example, adapts to the terrain. Then based on this network and the resulting parcel sizes, the population density and building density could be derived.

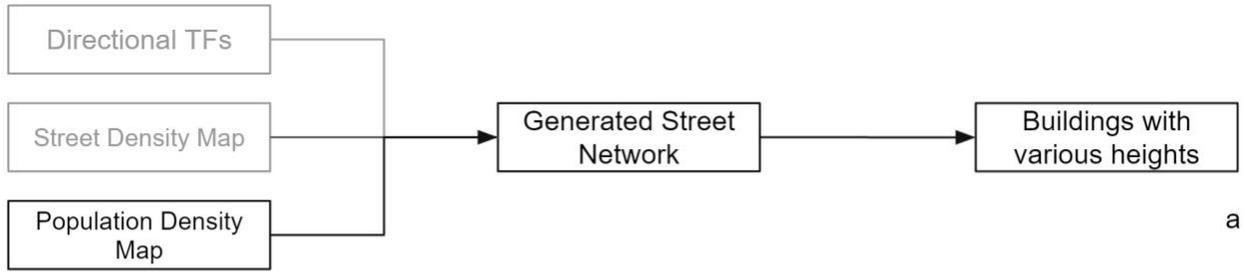

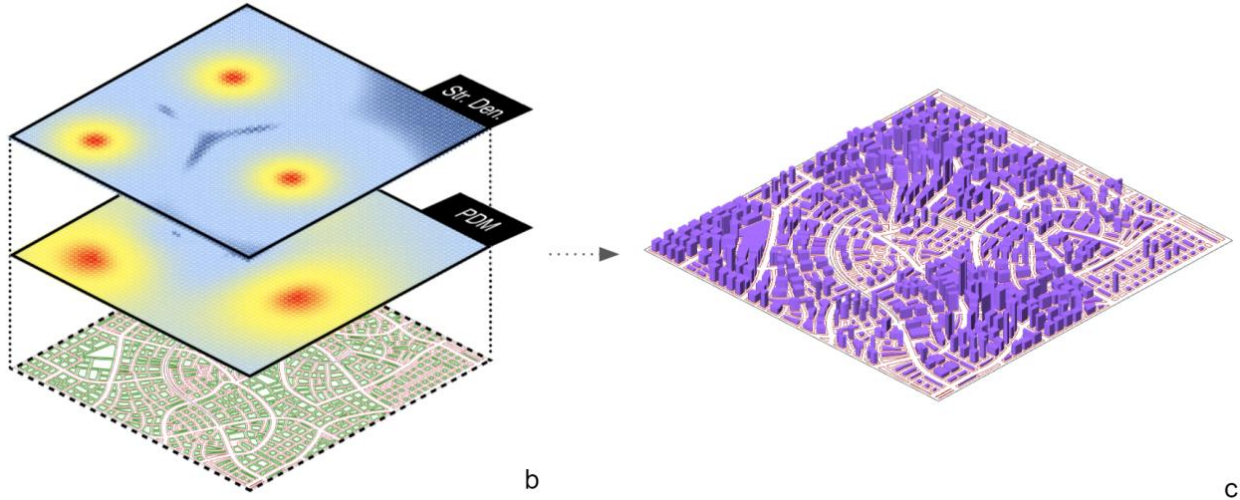

Figure 2. a) The flow of creating buildings of various heights. b) The input of PDM, which is coherent with the street density map before the generation of the street network. c) Generated 3D city.

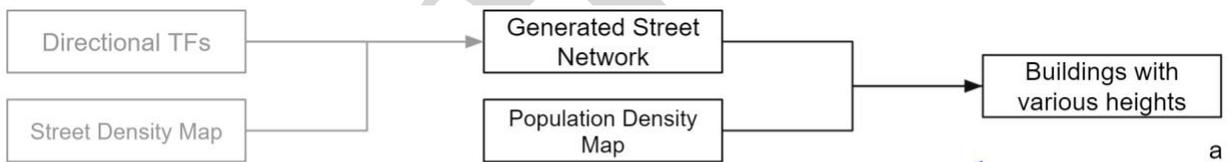

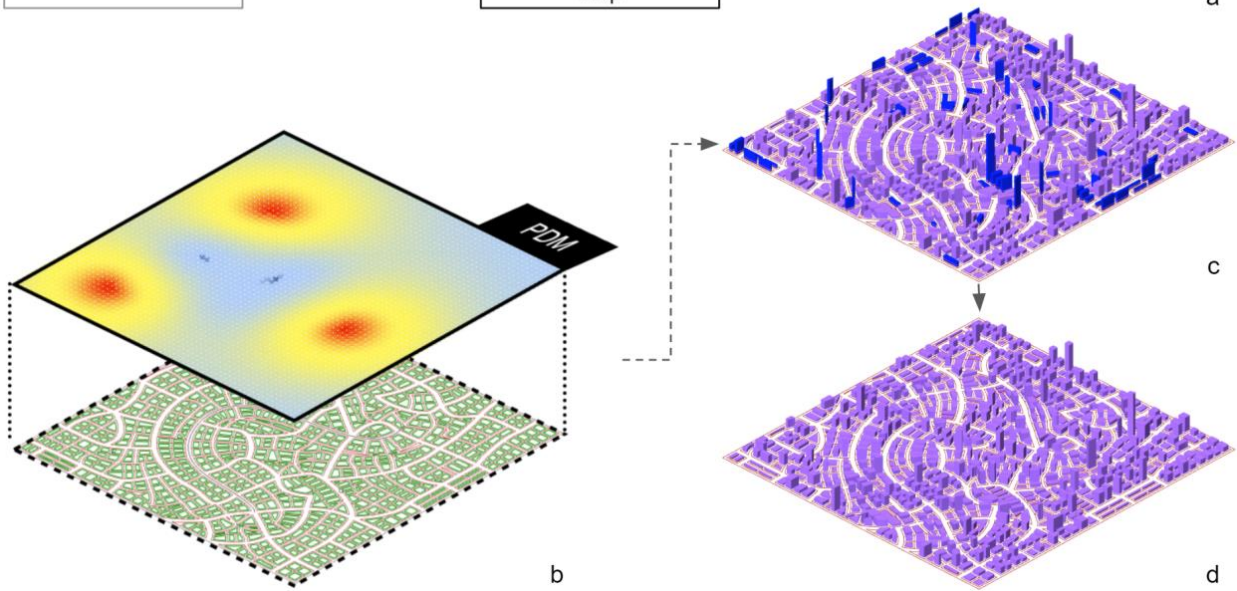

*Figure 3. a) The flow of creating buildings of various heights. b) The input of PDM, which is not necessarily consistent with the street density map, coupled with the generated network. c) Generated 3D city. d) The result after adding constraints of building area and aspect ratio.*

**Building Program Map (BPM):** Further, we showcase allocating amenities using desired building program (zoning) maps. Modelers can allocate building programs, amenities, and services within a city by providing a series of density maps. These maps can either be drawn or sketched by hand in a bitmap editor or generated by our framework using points or polygons. The BPMs are then superimposed, and building programs and amenities are then distributed in the model (Figure 4a). Note that magnitude in the map will be transferred as area ratio. Each pixel node will be assigned with an area ratio. Since a parcel can accommodate multiple nodes, we accumulate the area ratios within a parcel. For each type of land use, we specify an area target. Then, with the accrued area ratio, we can configure the floor area for the specific land use within the building parcel. Doing so for each type of land use inside the parcel and summing up the results will give us the total floor area for the parcel. Then divide the result with the lot's buildable area determines the building height. (Figure 4b).

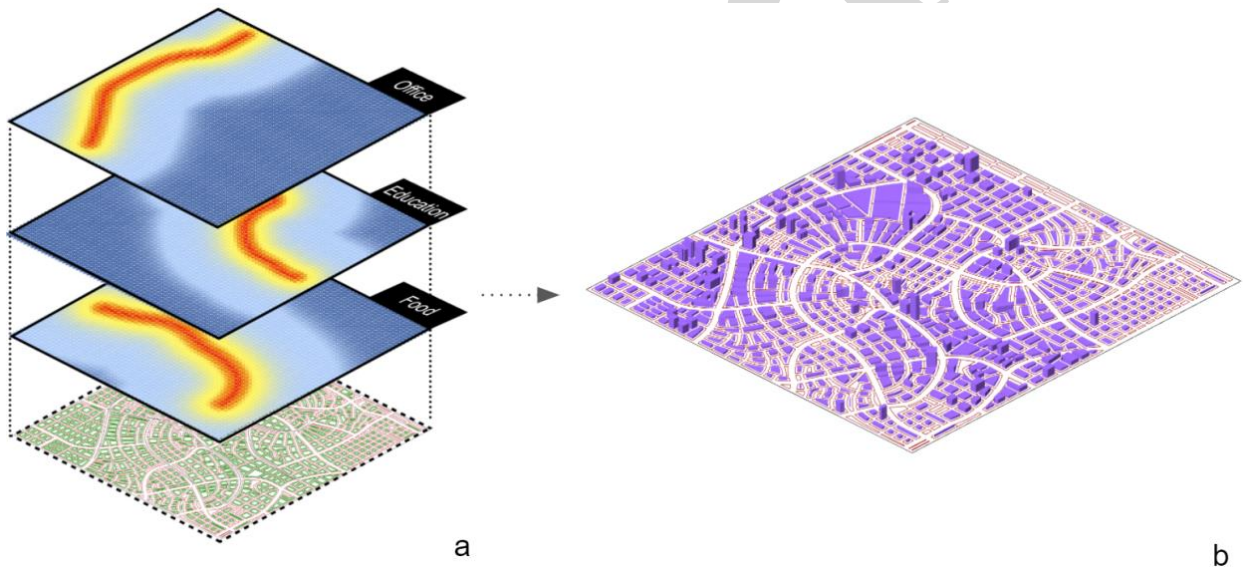

*Figure 4. a) Amenity strokes coupling with the generated street network. b) Generated 3D city.*

## Case Study

We present a simple case study as proof of our framework's concept. We use design space exploration to identify high-density, walkable design solutions that achieve good environmental performance.

### Metrics Setup

The goal of this exercise is to maximize urban density while offering good mobility and renewable energy potential. We assume there exists need for a workable tradeoff. For this case study, we select a series of metrics to quantify performance goals related to (1) urban density, (2) mobility, (3) energy demand, and (4) renewable energy potential.

**Density:** In this paper we consider population density map (PDM) to be a user input. However, the proposed methods that are based on tensor fields may also be used to derive PDM from urban data such as socio-demographic factors when available. All that would be required is to encode this into an image map which will be transferred to Magnitude Field in the proposed tool. Density in our exercise refers to the average floor area ratio, calculated by the division between the sum of the floor area of every building lot and the total buildable area within the design boundary, and population density that is measured as the fraction of total population and total buildable area. Note that all the design space exploration will be held within a predetermined fixed design boundary, therefore, as we increase the road density of a masterplan, the buildable area may decrease.

**Mobility:** We implement an active mobility analysis workflow proposed by (Yang et al., 2020), in which walkability ratings, commonly referred to as Walkscore (Dogan et al., 2018), consider population distribution and accessibility to amenities. With the input of BPMs, we distribute amenities and building populations to pixel nodes. After incorporating Urbano (Freeman, 1977), its mobility simulation function can infer a Walkscore to the corresponding masterplan scheme. During the process of solution space exploration, we will implement the rating for every scheme. Additionally, the betweenness centrality (Koenig et al., 2020) is another adopted metric and is computed with DecodingSpaces (Koenig et al., 2020). The metric in many applications is a proxy of congestion, thus will be minimized. It quantifies the number of times an intersection of a street network is acting as a bridge along the shortest path between two other intersections. So, there is a tradeoff between congestion and accessibility in our proposed scenario.

**Building Energy Demand:** To estimate potential building energy demand, we assume that all buildings meet an energy use intensity (EUI) target proclaimed by the AIA (Solemma, 2021). These EUI goals, which form the basis of the AIA 2030 Commitment, use CBECS 2003 (Commercial Building Energy Consumption Survey) and RECS 2001 (Residential Energy Consumption Survey) as the baseline EUIs and assume a 70% reduction in energy demand over the baseline. Different EUI targets are provided for different program types and are reprinted in Supp Fig S6. To estimate the total energy consumption, we simply aggregated AIA2030 EUI targets by the assigned building programs and corresponding floor. In future research, the models could also be coupled with dynamic energy simulations to provide higher fidelity energy consumption estimates.

**On-site Renewable Energy Potential (REP):** We compute the on-site renewable energy potential as the difference between the total amount of renewable energy produced from PV panels and the total building energy demand as estimated in the previous step. The power generated from PV panels depends on the usable area on facades and roofs of buildings. Climate Studio (Koenig et al., 2013) is used to compute the radiation levels and PV yield on the building envelope. In our simulation, the total usable area is assumed to be 40% of the façade area and 80% of the roof area. The PV panel efficiency is assumed to be at 20%. As a supplementary metric to REP, we also calculate the total surface area of building envelopes.

**Site:** The selected site (Supp Fig S7a) is in New Haven, Connecticut, where residential neighborhoods are on the south and the west, and high-density commercial districts are on the north and the east. Four accesses points were added to the design boundary. Since the site is adjacent to both residential and commercial context, we intend to generate a higher street density on the southwest (Supp Fig S7b, c), which then gradually drops towards the northeast (Supp Fig S7f, g). Additionally, four types of building program distribution maps and their respective proportions are specified in which 20% of land use is designated for offices (Supp Fig S7c), 60% for residentials (Supp Fig S7h), 5% for education facilities (Supp Fig S7i), and 15% for food and retail facilities (Supp Fig S7j). To calculate the buildable area of each plot, the floor area per person for each type of amenity is used. We assume an average floor area per person for housing domestic (36 $m^2$), office space required per person (8-13 $m^2$), for a full-service restaurant (2-3 $m^2$), and required space per student for an elementary school (9 $m^2$). Each type of building program map will lead to one buildable area value for one plot. In the case of plots with mixed usage, we simply sum up the buildable area for each program on the plot.

**Methodology:** We demonstrate that, by only altering two model parameters: The seed spacing, which governs the density of streets, and the population size, which controls the building massing, we can produce a wide range of diverse and plausible results. Additionally, we require the site to accommodate the target populations of 2k to 20k with an increment of 1k for each cluster of experiments. Fifteen seed spacing values are given within each cluster for the street generation.

The simulation results (Supp Fig S8) are visualized in Design Explorer (Hristov, 2021). Each polyline represents one solution in the ten-dimensional solution space (Figure 5). Seed spacing maintains a negative association with betweenness centrality. The trend of association between REP and betweenness centrality is not clear. A similar trend applies to REP and seed spacing as well. It results from that the value of building density interacts in a nonlinear fashion with the density of the street network. As the street-network becomes denser, the number of main and minor streets increases. This increases in the number of small parcels with solitaire buildings. This leads to more spacing between buildings and less overshadowing, better solar access, and higher PV yields. With the increasing population and resulting density, solar access is limited and, hence, denser and taller urban layouts are not always ideal solutions to achieving great REP (Figure 5a, b). This suggests that there is a density limit to achieving carbon neutrality. Using the computed data from design solutions, one can make a tradeoff between optimal objectives. We filter out designs with four metrics and conclude that under the given assumptions an ideal population size is around 9k to 11k (Supp Fig S9).

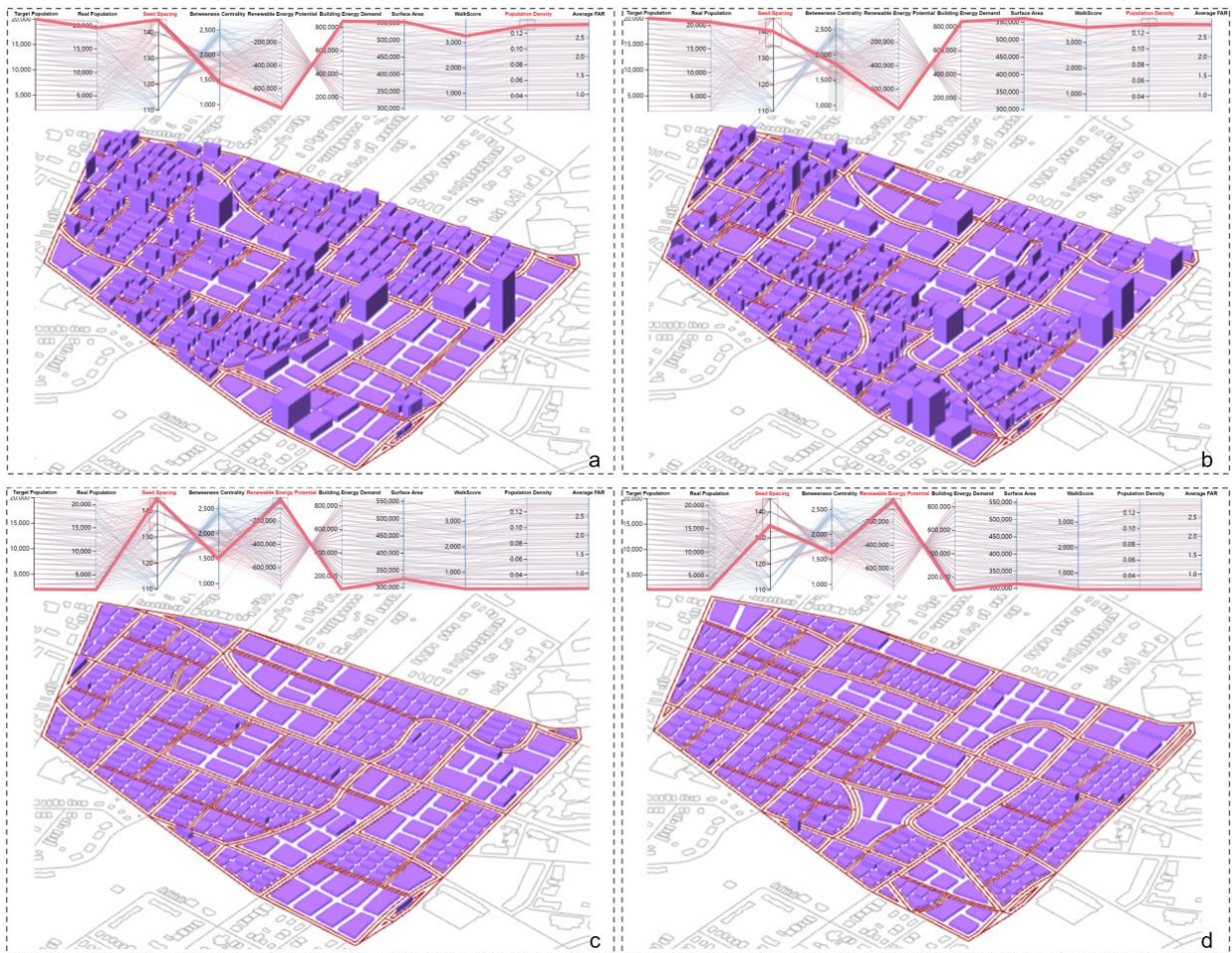

*Figure 5. Four instances of optimal solutions by selecting population density, seed spacing, and REP as metrics. The definition of the optimal solution and the value range of metrics for choosing optimal solutions are at the users' discretion. Figure a, b show denser and taller the layout of urban configuration is not always the ideal solution to getting great REP. Figure c, d indicate dense layouts with great REP are associated with low population density.*

## Discussion

The framework addresses common challenges in parametric urban design, including input complexity and high dimensionality of the design solution space. The challenge of generalizing various types of inputs asks modelers to transfer different design requirements into the same type of quantifiable problem representation for an algorithm. The translation hurdles are one reason why planners rarely embrace computer support (Miao et al., 2020; Wilson et al., 2019). Moreover, without a universal approach for automatic information translation, the resulting additional required work would not be feasible during the design process. The inefficiency potentially induces the limitation on how many parameters a designer can factor into the design in a short amount of time (Miao et al., 2020; Wilson et al., 2019). Therefore, our solution is designed to operate with high-level design inputs. As shown in case study, by modifying only two variables within a user-defined permissible range, the tool was able to generate a diverse range of design solutions. The high degree of automation can be attributed to the underlying tensor fields that provide users with a generalized way to encode contextual constraints and non-geometric design inputs with very few model inputs. Further, the idea of superimposing tensor fields helps to rationalize the complex logic and organization of urban fabric when we model cities as complex systems. It simplifies the modeling process by isolating various aspects of systems individually within corresponding tensor fields and then generates an urban fabric that takes their combined driving forces to generate urban form into account. The clear downside is that modelers only have indirect control over the design. However, if more direct control is needed, such as to force the generator to integrate specific design choices, like connections to existing streets or points of interest, modelers can specify local manipulations of the tensor field that can achieve the desired design outcome.

In the context of urban planning and design, it is common that designers need to deal with wicked problems elusive to a single correct answer as city morphologies are a by-product of many factors (Koenig et al., 2020). Therefore, starting a project from a thorough and complete problem description will be difficult (Koenig et al., 2020). Moreover, the diversity of stakeholders in an urban design process and their different views are also subject to change as the project evolves and thus further contribute to the uncertainty in the design process. Therefore, a framework that provides sufficient flexibility for users to add new or adjust existing aspects of the problem is much more meaningful than trying to find a single optimal solution. In our case study, with the combination of Design Explorer, we can identify the thresholds for each objective after the exploration of design space. The dimensionality of evaluation criteria can be easily adjusted at the user's discretion. The effects from the combination of preferences that are potentially contradictory to one and other (Koenig et al., 2020) are explicit during the tradeoff analysis. Such an assessment approach answers the need for nonlinear evaluation when dealing with various competing performance indicators, which would be difficult to realize in a classic iterative manual design process (Wilson et al., 2019). For instance, the volumetric design is devised upon the results of two-dimensional arrangement (street network and program layout). However, as the proposed case study shows, combining the two best results does not necessarily guarantee the best overall output. If aiming for the optimal REP, the dense street network is not the ideal foundation for the configuration of buildings under the condition of high population density (Figure 5a, b). This is difficult to realize in traditional design workflows, where the development of project is based on a more linear progression of design tasks that are not necessarily informed by feedback from simulation results and performance analysis of subsequent design phases.

It is the authors hope that the workflow's ability to generate urban 3d models quickly will facilitate the use of computational design space exploration coupled with simulation-based feedback. It is important to note that this implementation of the framework has limited modeling resolution when compared with the complexities that urban design must incorporate in practice. This may include detailed information on ownership of land, capacities and lanes in streets, turning paths of vehicles, and much more elaborate analysis of building typologies. The intentioned application of the proposed framework is primarily in early schematic design where the provided level of abstraction should be adequate to understand performance trade-offs of different designs. However, the authors hope that in future work, the proposed framework will be expanded to allow for even higher fidelity urban form generation and analysis.

## Conclusion

Design decisions of masterplan production such as zoning, program allocation, density and the layout of the street network, often involve careful consideration of plethora of information at the same time. To address the hurdle of generalizing different inputs, we have demonstrated a tensor-field-based framework for information interpretation and content generation. The entire proposed workflow has been detailed, including parsing data by directional maps and magnitude maps, generating street networks, parcellation and subdivision of blocks, and modeling buildings. The adaptivity and modularity illustrated in the methodology section provide designers with flexibility to produce diverse solutions. Additionally, we are aware that dealing with a complex system (i.e., a city) without systematically exploring an extensive range of urban forms that can be coupled with performance analysis potentially limit the quality of design results. Therefore, by incorporating several Rhino/Grasshopper-based urban analytical and environmental performance simulation tools, we showcased the feasibility of our framework to integrate multi-objective design performance analysis that can be used for trade-off assessment and optimization of masterplans. The proposed pipeline displays the capability to generate a diverse range of urban configurations with few inputs and exhibits the convenience to systematically search complex parametric design solution spaces. It promotes the efficiency of masterplan refinement at the early stage of the design process while keeping dimensionality and computational overhead manageable.

# Supplementary Material

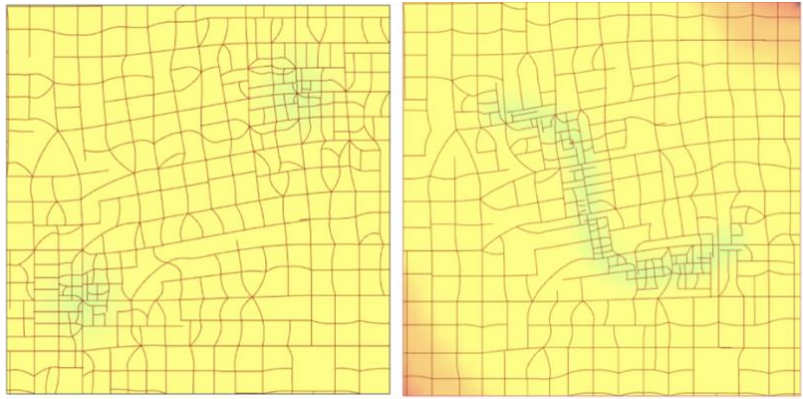

*Figure S1a. Density variation in one street network with magnitude field (e.g., population density map) as guidance.*

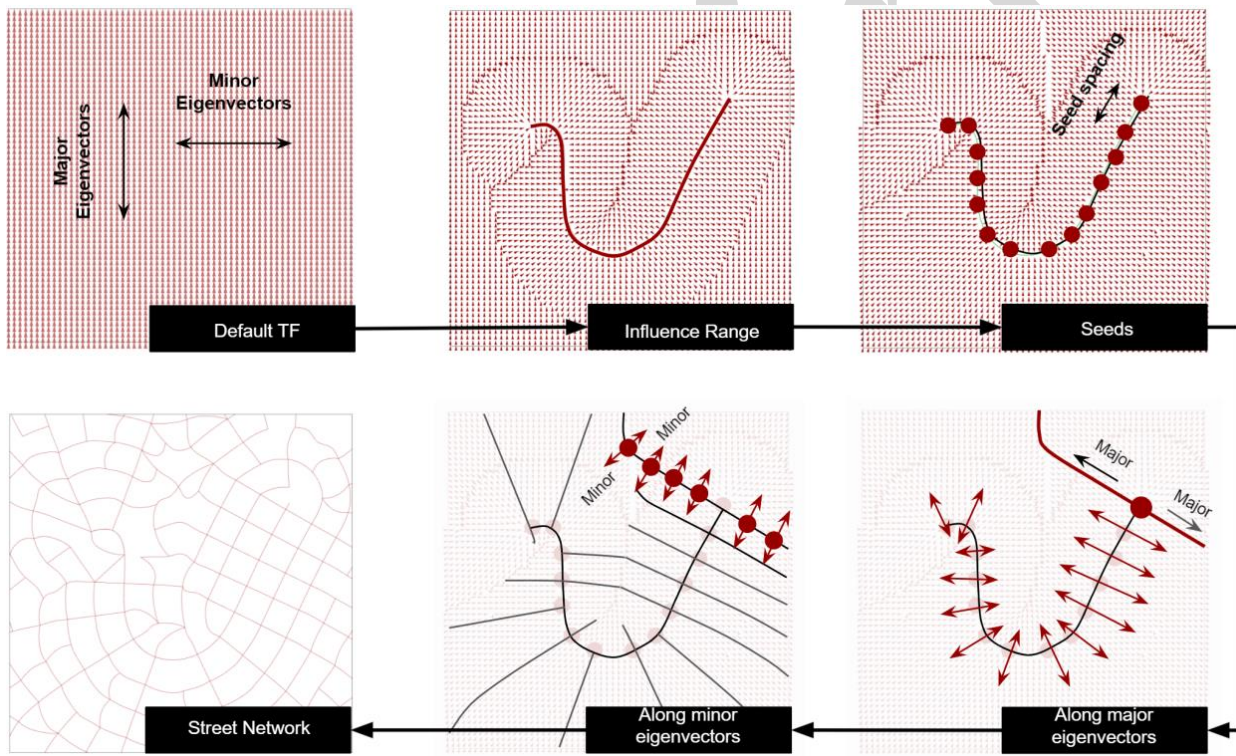

*Figure S1b. From tensor field to street network.*

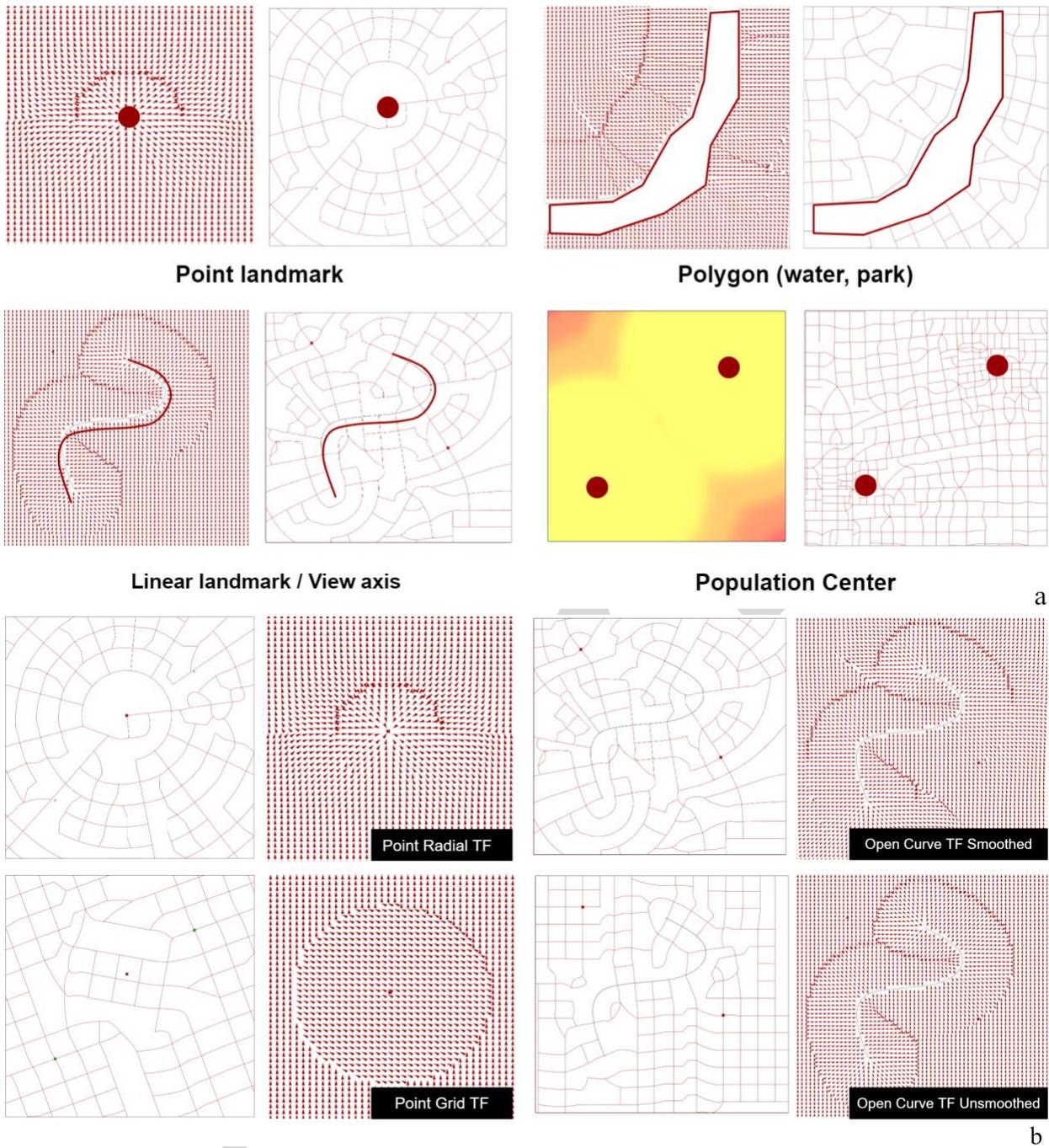

Figure S2. a) Tensor fields with different inputs and their corresponding street networks. b) Smoothed/unsmoothed TF and their corresponding networks.

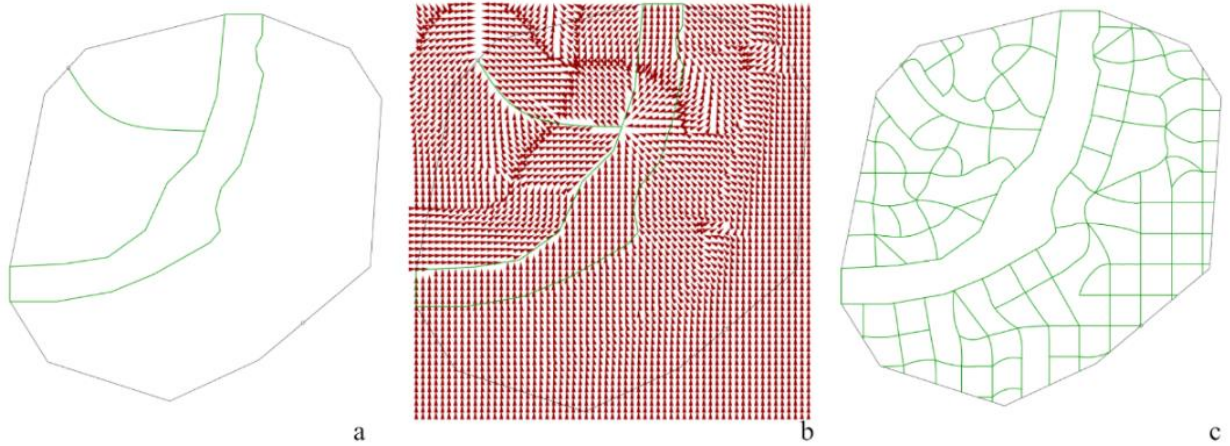

*Figure S3. Combination of two tensor fields (open curve tensor field and polygon tensor field).*

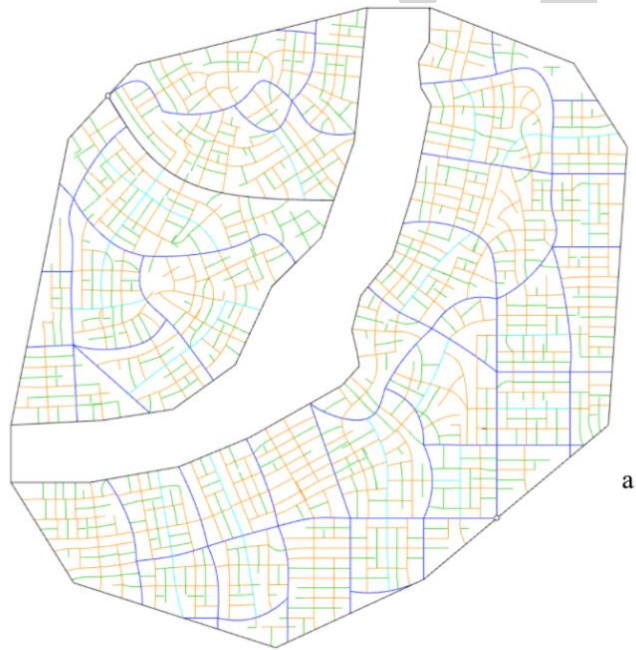

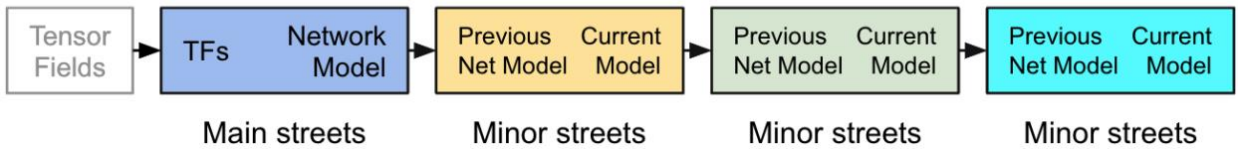

*Figure S4. Except for the generation of main streets, other levels of streets are produced by similar operations iteratively using the same component. a) 4 levels of the generated network. b) Chaining flow of streets generation.*

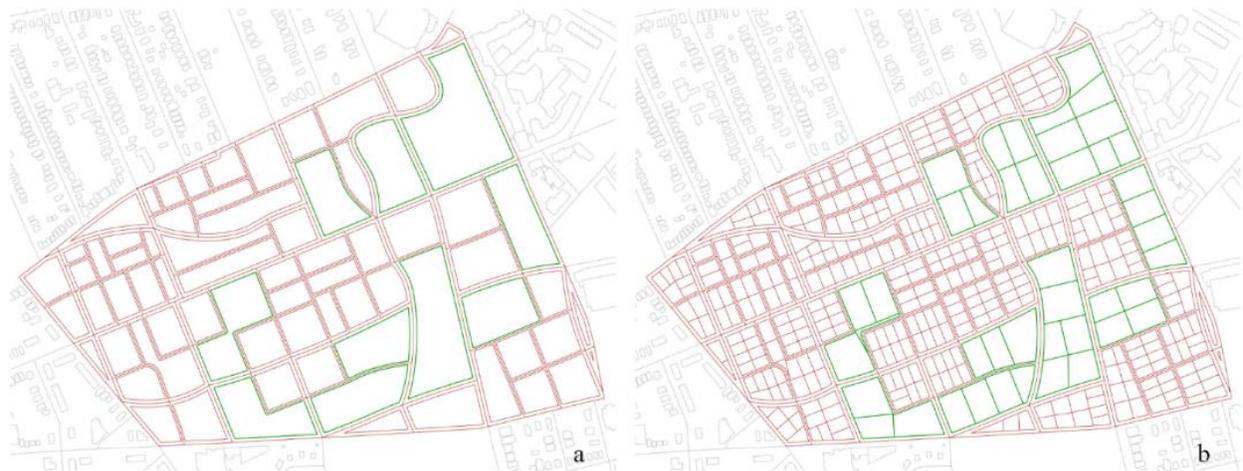

*Figure S5. Categorized subdivision so that building lots would not be partitioned to similar sizes. a) Parcels are separated into two groups. b) Subdivision result of different sizes of parcels.*

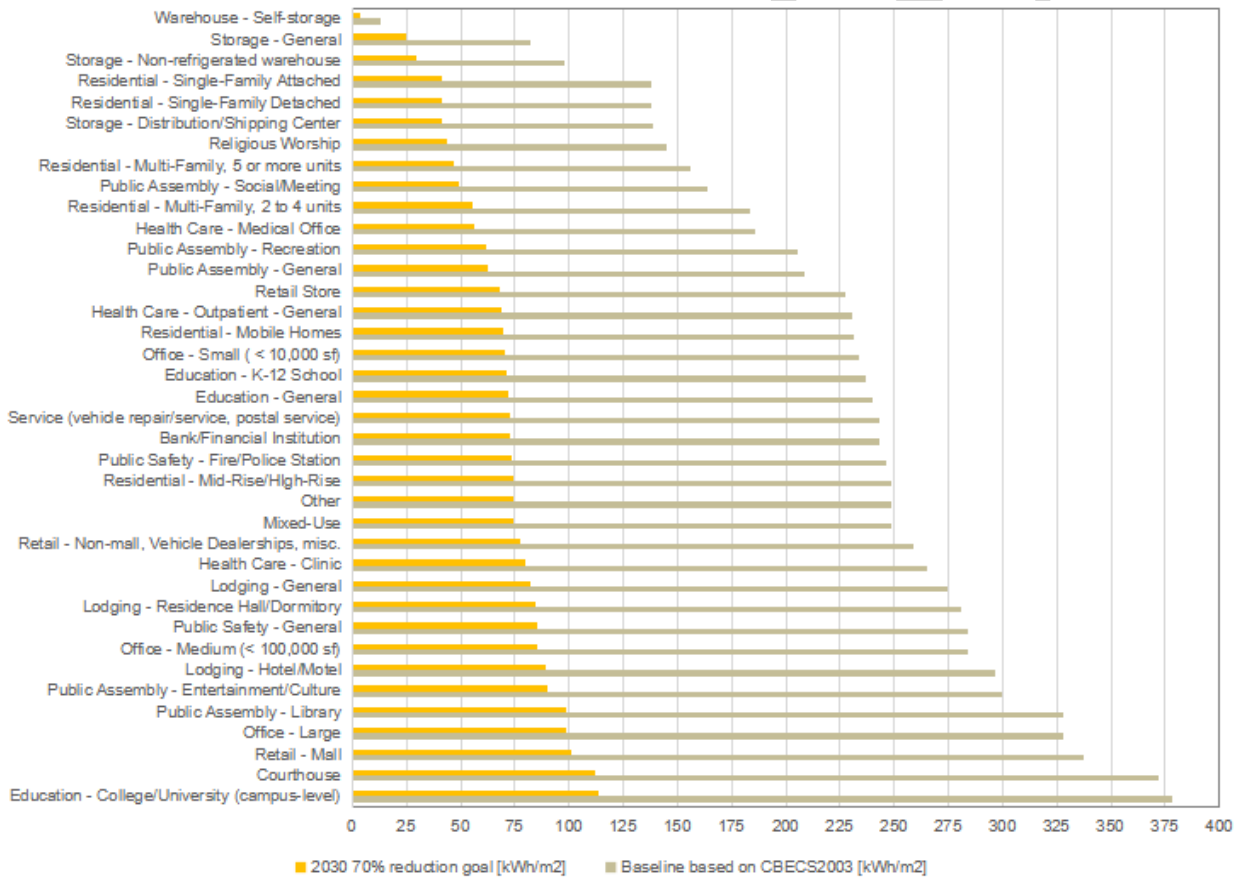

*Figure S6. AIA 2030 Baseline and Target EUI. To simplify our calculation, we only select four types of programs which are residential (mid/high rise), education (K12), office (medium) and retail store.*

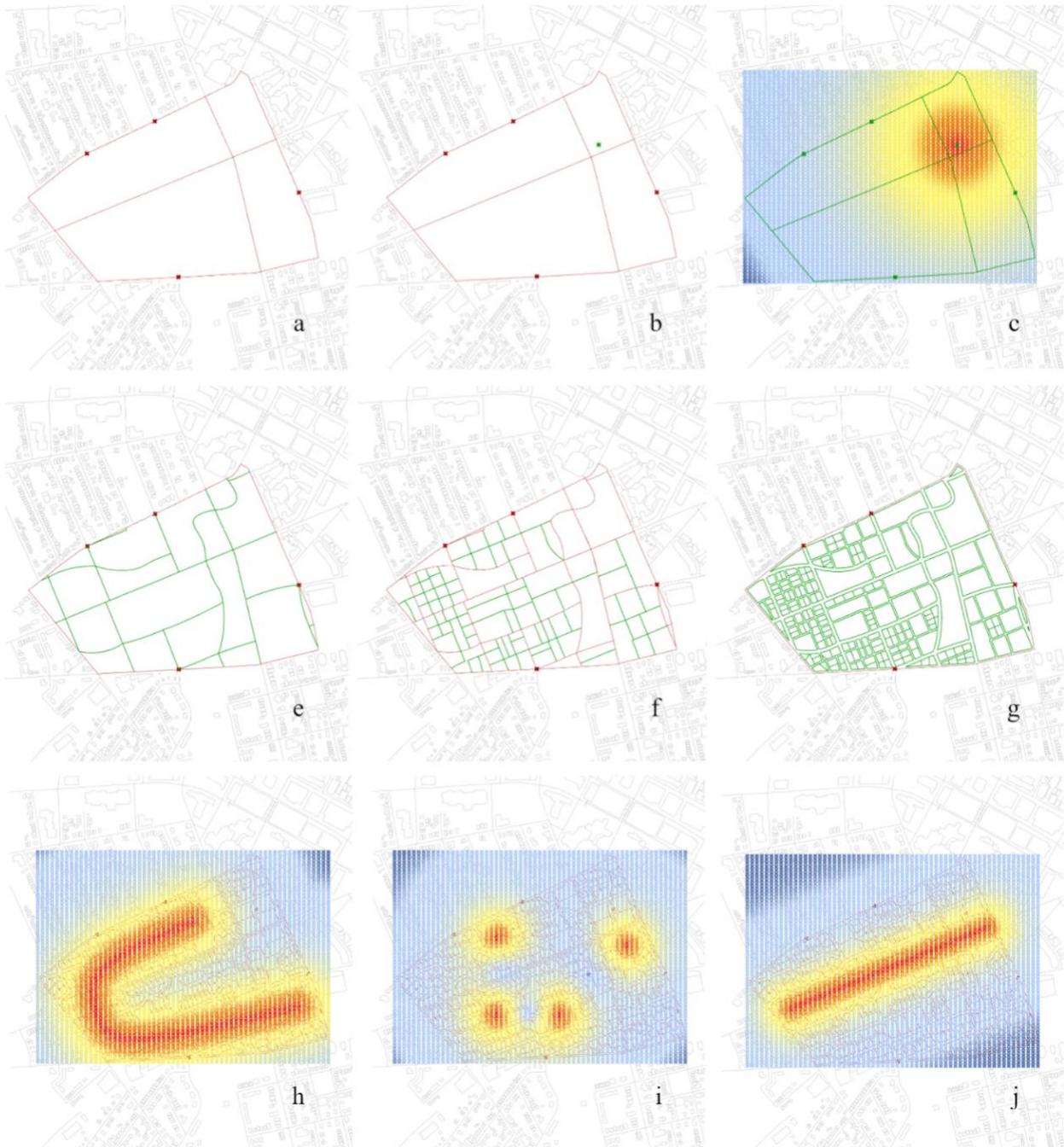

*Figure S7. The flow of the generation of the street network and parcellation. a) The site with additional accesses. b) The placement of the center of the street density map. c) The adjustment on street density distribution. e) The generation of main streets. f) The generation of the minor streets. g) The parcellation and subdivision based on the given density map. h) Residential BPM. i) Education BPM. j) Retail and Foodservice BPM.*

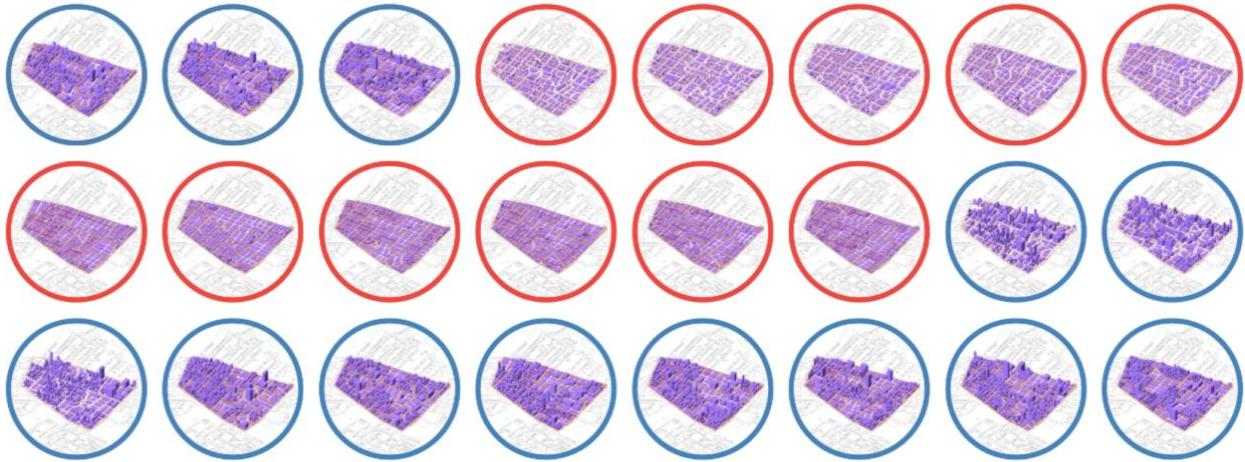

*Figure S8. Some instances of the solution space. The size of our design space is 152 solutions. It can be enlarged greatly depending on the number of manipulated parameters and the chosen ranges of them.*

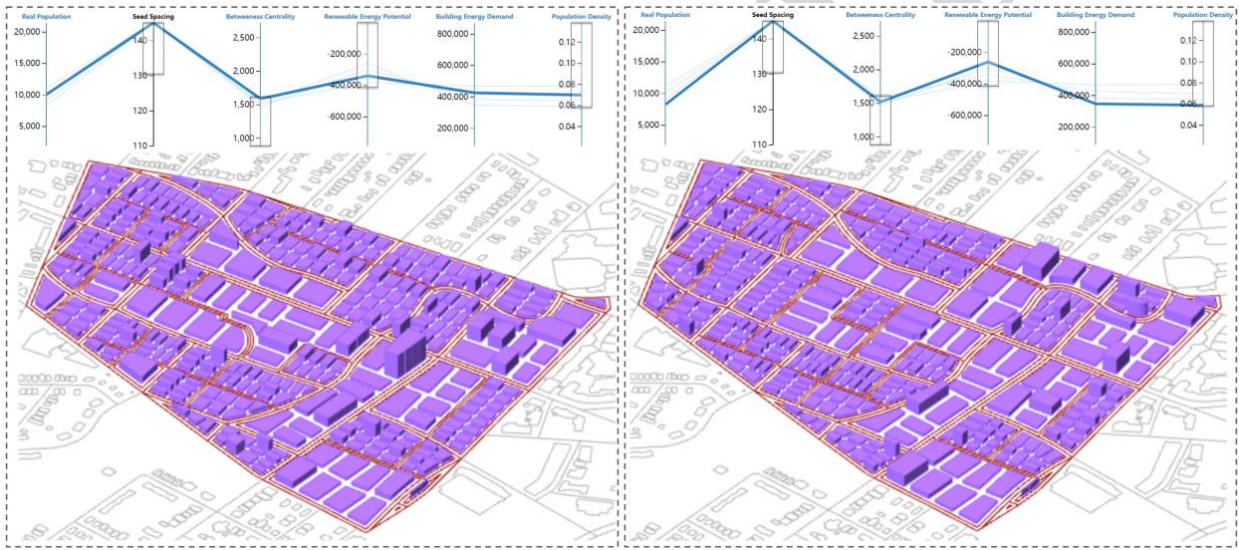

*Figure S9. A set of non-dominated configurations with relatively high average ratings by considering seed spacing, REP, betweenness centrality, and population density. We can infer from the figure the ideal range of population capacity within the given design boundary considering the chosen metrics.*